\documentclass[12pt]{article}

\usepackage[mathscr]{eucal}
\usepackage{amsmath,amsfonts,amssymb,amsthm}
\usepackage[hypertex]{hyperref}
\usepackage[matrix,arrow]{xy}

\theoremstyle{definition}

\newcommand{\bref}[1]{\textbf{\ref{#1}}}

\def\be{\begin{equation}}
\def\ee{\end{equation}}
\def\ba{\begin{array}}
\def\ea{\end{array}}
\def\d{\partial}
\def\dps{\displaystyle}

\newcommand{\half}{\frac{1}{2}}
\newcommand{\ads}{AdS_{2}}

\def\cA{\mathcal{A}}

\def\cM{\mathcal{M}}

\def\cR{\mathcal{R}}

\numberwithin{equation}{section} \makeatletter
\@addtoreset{equation}{section}

\begin{document}

\renewcommand{\thefootnote}{$\dag$}

\begin{flushright}
FIAN-TD-2013-10 \\

\end{flushright}

\vspace{8mm}

\begin{center}

{\Large\textbf{On higher spin extension of \\ \vspace{4mm} the Jackiw-Teitelboim gravity model }}

\vspace{.9cm}

{\large K.B.     Alkalaev \footnote{email: alkalaev@lpi.ru}}

\vspace{0.5cm}

\textit{I.E. Tamm Department of Theoretical Physics, \\P.N.
Lebedev Physical Institute,
\\ Leninsky ave. 53, 119991 Moscow, Russia}


\begin{abstract}

We formulate $AdS_2$ higher spin gravity  as  BF theory with
fields taking values in $sl(N, \mathbb{R})$ algebra treated as
higher spin algebra. The theory is topological and naturally
extends the Jackiw-Teitelboim gravity model so as to include higher spin
fields. The BF  equations linearized about $\ads$
background are interpreted as describing higher spin partially-massless fields of maximal depth
along with dilaton fields. It is shown that there are dual  metric-like
formulations following from the original linearized BF higher spin theory.
The duality establishes a dynamical equivalence  of the metric-like field equations that
can be given either as  massive scalar field equations or as
conservation conditions for higher spin currents.

\end{abstract}

\end{center}

\section{Introduction}

\renewcommand{\thefootnote}{\arabic{footnote}}
\setcounter{footnote}{0}

Two-dimensional gauge systems including generally covariant
theories and those with higher rank fields strike a balance
between descriptive  simplicity and nontrivial content.
Basically, this can be explained by that in two dimensions, unlike
in higher dimensions,  only scalar or spinor  modes can
propagate. It follows that some of gauge systems are topological
while others may exhibit  a quite complicated  dynamical structure which
encodes matter modes. Both cases are intrinsically related because
the matter can interact via topological gauge fields, while matter
modes themselves can be represented as particular components of
gauge fields. This is  most clearly illustrated by the
massive  Schwinger model of fermions interacting via topological
Maxwell field, or by the Liouville field originating as the metric
component in the conformal gauge.

A prominent example of topological gauge  systems is the Jackiw-Teitelboim gravity model
\cite{Teitelboim:1983ux,Jackiw:2D} which replaces the  standard gravitation theory. Recall that in two dimensions
Einstein tensor vanishes identically so that the simplest alternative  is given by  diffeomorphism invariant
equation of motion of the form \footnote{Manifold $\cM^2$ is a general two-dimensional space-time
with local coordinates $x^m$,  Lorentz world indices run $m,n = 0,1$, Lorentz fiber
indices run  $a,b = 0,1$, $o(2,1)$ fiber  indices  run
$A,B,C = 0,1,2$, $o(2,1)$ invariant metric is $\eta^{AB} = (+--)$.
The Levi-Civita tensor $\epsilon_{ABC}$ is normalized as $\epsilon_{012}=+1$.
Two-dimensional anti-de Sitter spacetime $AdS_2$ has a radius $L$
and a signature $(+-)$, so that the cosmological constant is
$\Lambda = -1/L^2$. The Levi-Civita tensor $\epsilon_{mn}$ is normalized as $\epsilon_{01} = +1$.}
\be
\label{JTeq}
R-\Lambda = 0\;,
\ee
where $R$ is the Ricci scalar curvature built of the metric field
$g_{mn}(x)$, while $\Lambda$ is the cosmological constant. \footnote{The Riemann tensor in two dimensions has only
a scalar curvature component  as one can see from the identity
$\epsilon^{mn}\epsilon^{kl}R_{mn,kl} = 2 R$.}
The above theory is obviously non-Lagrangian since one has a single equation for three variables.  Adding  an auxiliary scalar field $\phi(x)$
allows one to introduce  the variational principle so that  the
resulting local action describes a particular dilaton gravity. The
theory is however topological and contains  no local degrees of
freedom. At the same time, $\ads$ spacetime and  black hole
type objects are  solutions of the Jackiw-Teitelboim  model
\cite{Cadoni:1994uf,Spradlin:1999bn}.

The main purpose  of this paper is to elaborate a higher spin generalization
of pure gravity system governed by  equation \eqref{JTeq}. We introduce higher rank gauge fields and their
equations of motion in such a way that the resulting gauge system contains the
gravitational subsector. As a guiding principle, we
employ the frame-like formulation instead of the metric-like one
used to postulate  equation \eqref{JTeq}. The idea behind the frame-like approach
originates from the Cartan formulation of the Einstein  gravity where
the metric is the symmetric component of the vielbein, while the
antisymmetric component is a pure gauge by local Lorentz
transformations.

Indeed, when extending to higher spins,  the Jackiw-Teitelboim model viewed as a gauge theory of
the $\ads$ global symmetry algebra $o(2,1)\sim sl(2, \mathbb{R})$
with a BF local action
\cite{Fukuyama:1985gg} has
conceptual advantages over alternative approaches. One  considers  BF action with fields
taking values in $sl(N, \mathbb{R})$ algebra with a finite  parameter $N$.
We show that such a theory can be
interpreted  as two-dimensional topological  $\ads$ higher spin gravity
theory. Note that conventional higher-spin metric-like fields $g_{m_1 ... m_s}$ with  $s$ Lorentz  indices
appear as particular tensor components of $sl(N, \mathbb{R})$ gauge connection
rewritten in $o(1,1) \subset o(2,1) \subset sl(N, \mathbb{R})$ manifestly Lorentz covariant basis.
Other components are either auxiliary or pure gauges.
We find that the set  of local fields  over $\ads$ background consists of
gauge fields with spins $s = 2,3, ... , N$ and masses $m^2_s =
s(s-1)\Lambda$ interpreted as topological  partially-massless
fields of  maximal depth. \footnote{The dilaton sector comprising
$0$-form fields of the BF formulation will be analyzed elsewhere  \cite{Alkalaev_soon}.}

On going from the original   BF formulation linearized over  $\ads$ space
to the metric-like formulation one finds out that there are two different but dynamically equivalent forms
of the metric-like theory. In particular, it turns out that the gauge sector of the quadratic theory
can be
dually represented either by massive scalar equations or by conservation conditions for higher spin currents.
Both scalar/current equations are invariant with respect to  on-shell
symmetries/impro\-ve\-ments that guarantee the absence of local degrees of freedom.

While the BF theories under consideration are topological, we have in mind to study $\ads$ models
with local degrees of freedom
interacting via higher spin topological fields. Such an interacting theory
will  contain at least cubic field-current interactions of the form
$\sum_{s} \int d^2 x \sqrt{g_{_{AdS_2}}}\;\, \Omega_m{}^{a_1 ... a_s} J^m{}_{a_1 ... a_s}$,
where $\Omega_m$ are some Lorentz spin-$s$ gauge fields, while $J_m$ are conserved Lorentz spin-$s$ currents built
of matter fields. \footnote{ A particular example of such a   $d=2$ higher spin model has been proposed  by Vasiliev \cite{Vasiliev:1995sv}.
Remarkably, it  is formulated as BF theory with fields taking values in some infinite-dimensional algebra,
and up to now this is the only example of fully  non-linear higher spin gauge theory with local degrees of freedom
formulated at the action level.} We suggest that the field-current duality mentioned  above will play an important role
in the analysis of the higher spin field-current couplings in $\ads$ space. It follows that one of the aims of the present paper is to
classify possible higher spin gauge fields in $\ads$ space
that may couple to conserved higher spin currents.

\section{The BF higher spin extension}
\label{sec:higherspinext}

Gravitational fields in two dimensions can be  identified with  $o(2,1)$ connection
$W^A(x)\, T_A =  d x^m\, W^A_m(x)\, T_A$, where $T_A$ are $o(2,1)$ basis elements with the commutation relations
$[T_A, T_B] = - \epsilon_{ABC}\,T^C$. \footnote{Note that the basis elements can be equivalently
represented in the dualized form as  $T^{A} = \frac{1}{2}  \epsilon^{ABC} T_{BC}$. The commutation
relations for antisymmetric basis elements $T^{AB} = -T^{BA}$ take the standard form $[T^{AB}, T^{CD}] = \eta^{AD}T^{BC} + \text{3 terms} $.}
Adding $0$-form fields $\Psi(x)  = \Psi^A(x)\, T_A$ one arrives at the BF action \cite{Fukuyama:1985gg}
\be
\label{JTaction}
S_{JT}\,[W,\Psi] = \int_{\cM^2} \Psi_A  \cR^A\;,
\ee
where the curvature is given by $\half dx^m \wedge dx^n \,\cR_{mn}^A \equiv \cR^A =  d W^A - \epsilon^{ABC} W_B \wedge W_C$.
The equations of motion following  from  \eqref{JTaction} are
\be
\label{zerc}
\cR_{mn}^A = 0\;,
\quad \qquad
D_m \Psi^A = 0\;,
\ee
where $D_m = \d_m + W_m$ is the  $o(2,1)$ covariant derivative.
The 1-form connection denoted as $W^A_0 = \big(\sqrt{-\Lambda}\;h^a_m,\, w_m\big)$ corresponds  to
$AdS_2$ spacetime. The zero-curvature constraint $\cR_{mn}^A(W_0) = 0$ expresses Lorentz spin connection $w_m$
via the zweibein  $h^a_m$, while the latter yields  $AdS_2$ spacetime metric $g_{mn}$ through the standard identification
$g_{mn} = \eta_{ab} h^a_m h^b_n$, where the fiber Minkowski metric is $\eta_{ab} = (+-)$.

The BF  frame-like
equations of motion are dynamically equivalent to the metric-like
equations  of original Jackiw-Teitelboim model. In particular, the original equation
\eqref{JTeq} is identified with one component of the curvature  \eqref{zerc}, while other two components
are constraints  (see \cite{Fukuyama:1985gg} for more details).
Despite the  general covariance, the theory describes
a scalar Liouville mode  that follows from
equation \eqref{JTeq} in the conformal gauge. However, the theory is topological
because the resulting Liouville equation possesses a residual coordinate invariance that gauges away
all functional freedom of the solutions, see, \textit{e.g.,} \cite{Jackiw:1992bw}.

The higher spin generalization of the theory \eqref{JTaction} is straightforward. Two-dimensional BF theory with $\cA$-valued fields, where $\cA = sl(N, \mathbb{R})$
can be considered  as particular higher spin gauge theory.
In order to introduce higher spin gauge fields  one decomposes the adjoint of $sl(N, \mathbb{R})$
algebra into totally symmetric irreps of $AdS_2$ spacetime global symmetry algebra
$sl(2, \mathbb{R}) \approx o(2,1)$. For instance, in the case of $\cA = sl(3, \mathbb{R})$
the  basis elements $T^\alpha$  labelled by  $\alpha = 1,..., 8$
can be equally rearranged  as $T^A \oplus T^{(AB)}$ with $A,B =
0,1,2$, \textit{i.e.},  as a direct sum of rank-$1$ and rank-$2$
totally symmetric and traceless $sl(2, \mathbb{R})$ algebra tensors. In
particular, respective one-form connections are  identified
with frame-like fields of spins $2$ and $3$.
The analogous construction has been used  in
the context of $3d$ Chern-Simons  higher spin theory with algebra
$\cA\oplus \cA$, where
$\cA= sl(3, \mathbb{R})$ \cite{Henneaux:2010xg,Campoleoni:2010zq}.

It follows that Lie algebra $sl(N, \mathbb{R})$ for $N = 2,3,...$ can be interpreted as a finite-dimensional higher spin
algebra in two dimensions provided that its basis elements are organized as follows
\be
\label{sln}
T_{A_1}\, \oplus\, T_{A_1A_2}\, \oplus\; \cdots \; \oplus\, T_{A_1 ... A_{N-1}}\;,
\ee
where $T_{A_1 ... A_k}$ are rank-$k$ totally symmetric and traceless $sl(2, \mathbb{R})$ algebra tensors,
\be
\label{irrgen}
T_{(A_1 ... A_k)}\;:
\qquad
\eta^{MN} T_{MNA_3 ... A_k} = 0\;.
\ee
It is worth noting that the decomposition \eqref{sln} comes from   the principal  embedding of
$sl(2, \mathbb{R}) \subset sl(N, \mathbb{R})$.  \footnote{A non-principal embedding gives rise
to a different dynamical content, and, generally speaking, to a different theory. In the context of
$3d$ Chern-Simons higher spin theory non-principal embeddings and respective spectra were discussed, \textit{e.g.,}  in \cite{Castro:2012bc}.}

Introduce differential 0-form and 1-form fields taking values in $\cA = sl(N, \mathbb{R}) $
with basis elements \eqref{sln}
\be
\label{connec}
\Psi(x)  = \sum_{s=2}^{N-1}\Psi^{A_1 ... A_{s-1}}(x)\, T_{A_1 ... A_{s-1}} \;,
\qquad
W(x) = \sum_{s=2}^{N-1}\,dx^m\, W_m^{A_1 ... A_{s-1}}(x)\, T_{A_1 ... A_{s-1}} \;.
\ee
Expansion coefficients in the  higher spin algebra basis elements  \eqref{sln} are to be identified
as higher spin fields.  They satisfy $o(2,1)$ irreducibility conditions of the form \eqref{irrgen}.

Let us define  $\ads$ higher spin gravity by the following
BF action
\be
\label{action}
S[ W, \Psi] = g \sum_{s=2}^{N-1} \,\int_{\cM^2}   \Psi_{A_1 ... A_{s-1}} \cR^{A_1 ... A_{s-1}} \;,
\ee
with  curvature two-forms   given by
$\cR_{mn}^{A_1 ... A_{s-1}} = \d_{[m}^{} W^{A_1 ... A_{s-1}}_{n]} + [W_m, W_n]^{A_1 ... A_{s-1}}$,
where $[\cdot, \cdot]$ is the Lie bracket  in $\cA$ evaluated in basis \eqref{sln},
and $g$ is a dimensionless coupling constant. The action  \eqref{action} is invariant under the  $sl(N, \mathbb{R}) $ gauge  transformations
\be
\label{fulgaugetr}
\delta W_m^{A_1 ... A_{s-1}}= \d_m \xi^{A_1 ... A_{s-1}} + [\xi,W_m]^{A_1 ... A_{s-1}}\;,
\qquad
\delta \Psi^{A_1 ... A_{s-1}} =  [\xi, \Psi]^{A_1 ... A_{s-1}}\;,
\ee
where $\dps\xi(x) = \sum_{s=2}^{N-1}\xi^{A_1 ... A_{s-1}}(x) T_{A_1 ... A_{s-1}}$ is an
$\cA = sl(N, \mathbb{R})$-valued gauge parameter.
Note that \textit{a-priori} there is no  metric tensor  as the theory is completely formulated via differential
forms. It follows that the theory is diffeomorphism invariant. However, an on-shell diffeomorphism
transformation is simply a gauge transformation with particular gauge parameter.

Variational equations of motion for $\Psi^{A_1 ... A_{s-1}}$ and $W_m^{A_1 ... A_{s-1}}$  are
the zero-curvature equation and the covariance constancy equation
\be
\label{curveq}
\cR_{mn}^{A_1 ... A_{s-1}}= 0\;,
\qquad\;\;
D_m \Psi^{A_1 ... A_{s-1}} = 0\;, \qquad\;\;   s = 2,..., N-1\;,
\ee
where $D_m =\d_m + W_m$ is a covariant derivative with 1-form connection $W_m$ given by \eqref{connec}.
The covariance constancy equation involves both  0-form and $1$-form fields, while the zero-curvature
equation forms an independent gauge sector of the theory that can be analyzed separately.
Let us emphasize that solutions of the zero-curvature equation  are  given locally by pure gauge 1-form fields.
For the case of finitely many gauge fields this observation guarantees that the theory is topological and therefore
contains no local  degrees of freedom. However, in the next
section we explicitly examine the gauge sector in the free field approximation to provide a particular
physical interpretation of  topological modes.

\section{Linearized higher spin dynamics in $AdS_2$ space}
\label{sec:lindyn}

The higher spin theory \eqref{action} contains a gravitational  sector described by
the Jackiw-Teitelboim action. The gravitational fields are singled out by setting all
higher spin fields to zero except for those taking values in $sl(2,
\mathbb{R})$ subalgebra. This particular truncation is obviously consistent. At the same time,
$\ads$ space is the ground state of the theory since it solves gravitational equations \eqref{JTeq}.

Since the action \eqref{action} is non-linear containing at most cubic terms $\sim \Psi W^2$, one can consider the perturbation theory over a particular
solution. To this end introduce   1-form  $W_0$ describing $AdS_2$ spacetime and
0-form  $\Psi_0 = 0$ as the background of the theory, so that
all fields are decomposed as $W = W_0 + \Omega$ and $\Psi = \Psi_0
+ \Phi$, where $\Omega$ and $\Phi$ are fluctuations over the
background. The background fields $W_0$ and $\Psi_0$ obviously
solve BF higher spin  equations \eqref{curveq}. Similarly, the 2-form curvatures are expanded as $\cR = \cR_0+R+ ...$,
where $\cR_0  \equiv \cR(W_0)$ is the zero-curvature gravitational
constraint \eqref{zerc}, while $R$ is the linearized curvature and
the ellipsis denote second order corrections.

In the one-form sector, gauge fields $\Omega$ and their linearized curvatures $R$ have the component form
\be
\label{gaugefield}
\Omega^{A_1 ... \, A_{s-1}} = dx^{m}\, \Omega_{m}{}^{A_1 ... \, A_{s-1}}\;,
\qquad
R^{A_1 ... \, A_{s-1}}  = D_0\Omega^{A_1 ... \, A_{s-1}}\;,
\ee
where $D_0 = d + W_0$ is the background covariant derivative.
Representing  the zero-curvature condition \eqref{zerc} as $\cR(W_0) \equiv  D^2_0 = 0$ one obtains
that linearized  curvatures are invariant under the gauge transformations \eqref{fulgaugetr} expanded over the background,
\be
\label{trans}
\delta \Omega^{A_1 ... \, A_{s-1}} = D_0 \xi^{A_1 ... \, A_{s-1}}\;,
\ee
where $\xi^{A_1 ... \, A_{s-1}}$ are 0-form gauge parameters. The Bianchi identities
$D_0 R^{A_1 ... \, A_{s-1}} \equiv 0$  following from  \eqref{gaugefield} are trivial since any
3-form in two dimensions  vanishes identically. Fluctuations in the zero-form sector
are inert under the gauge transformations
$\delta \Psi^{A_1 ... A_{s-1}} = 0$, at least to the lowest order in the fields, cf.
\eqref{fulgaugetr}.

Let us compare the resulting set of 1-form higher spin gauge fields \eqref{gaugefield} in $d=2$ dimensions
with fields   in  $d\geq 3$ dimensions. By massless higher spin fields in $AdS_{d}$ we call 1-form
connections $\Omega^{A_1 ... A_{s-1},\, B_1 ... B_{s-1}}_m$ taking values in $o(2,d-1)$ finite-dimensional irreps
of the symmetry type described by two-row rectangular Young diagram \cite{Vasiliev:2001wa}. In the metric-like
formulation these fields reduce to Fronsdal spin-$s$ massless fields. Obviously,
for $d  = 2$ all 1-form connections vanish identically except for $s=2$ case of the gravity, $\Omega_m^{A_1,\, B_1} = -
\Omega_m^{B_1, \,A_1}$. It follows  that there are no spin $s>2$ Fronsdal fields  in $\ads$.

At the same time, in two
dimensions fiber indices of  spin-$2$ connection can be dualized by virtue of $o(2,1)$ Levi-Civita tensor
as $\Omega_m^{A_1} = \epsilon^{A_1MN}\Omega_m{}_{M,\,N}$.
The point is that the gravitational connection $\Omega_m^{A}$ starts  an infinite
sequence  of  1-form connections $\Omega^{A_1 ...\, A_{s-1}}_m$, $s=2,3,... $, taking values in $o(2,1)$
finite-dimensional irreps
described by one-row Young diagrams. These are fields appearing  via the gauging
of $sl(N, \mathbb{R})$ algebra in the basis \eqref{sln}. Note that in the $d \geq 3$ case,
the connections with totally symmetric $o(d-1,2)$ fiber indices  describe partially-massless
fields of maximal depth \cite{Deser:2001us,Zinoviev,Skvortsov:2006at}. It follows that in $\ads$ space
all possible higher spin $s>2$ fields are exhausted by "topological  partially-massless" fields of
maximal depth. \footnote{In higher dimensions, the depth is a number of derivatives in the gauge transformation law
for a spin-$s$ dynamical field, $t = 1,...,s$. In two dimensions, a  maximal value of the depth equals $t=s-1$
due to the Hodge duality mentioned above.}


The BF equations \eqref{curveq} linearized around $\ads$ background space
decouple into a direct sum of $N-1$ independent subsectors
\be
\label{1eq}
D_0 \Omega^{A_1 ... A_{s-1}} = 0\;,
\ee
\be
\label{2eq}
D_0 \Phi^{A_1 ... A_{s-1}} = 0\;,
\ee
each of spin $s = 2,3,..., N$. In what follows, we analyze a set
of topological metric-like fields encoded in the gauge sector described by equations \eqref{1eq}.

To identify  the  dynamical content of equations \eqref{1eq} one decomposes $o(2,1)$ covariant gauge fields
and linearized curvatures \eqref{gaugefield}  into Lorentz algebra $o(1,1) \subset o(2,1)$  irreducible components,
\be
\label{dectraceless}
\Omega_{m}^{A_1 ... \, A_{s-1}} = \bigoplus_{k=0}^{s-1}\; \omega_{m}^{a_1 ... \, a_k}\;,
\qquad
R_{mn}^{A_1 ... \, A_{s-1}} = \bigoplus_{k=0}^{s-1}\; R_{mn}^{a_1 ... \, a_k}\;,
\ee
\textit{i.e.}, totally symmetric and traceless with respect to the Minkowski tensor $\eta^{ab}$.
Assuming that Lorentz indices $a_i$ are symmetrized with a unit weight,
we find that the component form of \eqref{gaugefield} is given by
\be
\label{curva}
\ba{l}
\dps
R_{mn}^{a_1 ... a_k} = \nabla^{}_{[m} \, \omega_{n]}^{a_1 ... a_k}
-\Lambda \frac{(s-k-1)(s+k)}{2(k+1)} h^{}_{[m}{}_{,\,c} \wedge\omega_{n]}^{ca_1 ... a_k} +
\\
\\
\dps
\hspace{6cm}+\Big[h_{[m}^{a_1}\, \wedge \omega_{n]}^{a_2 ... a_k}  - \frac{1}{k-1}\, \eta_{}^{a_1a_2} h^{}_{[m,\,c}\, \wedge\omega_{n]}^{c a_3 ... a_k}\Big]
\;,
\ea
\ee
where  covariant derivative $\nabla_{m} = \d_m + w_m$ is defined with respect to the background
Lorentz spin connection $w_m$, while $h^a_{m}$ is the background zweibein.
Quite analogously, one finds  the component form of the gauge transformations \eqref{trans},
where gauge parameters $\xi^{a_1 ... a_k}$ are $o(1,1)$ irreducible components
of the $o(2,1)$ gauge parameter $\xi^{A_1 ... A_{s-1}}$ \eqref{trans}.

The Lorentz component form of BF equations \eqref{1eq}  is
\be
\label{main}
R^{\,a_1 ... \,a_k}_{mn} = 0 \;,
\qquad
k = 0, 1, ..., s-1\;.
\ee
In the  spin $s=1$ case the only field $\omega_m$ is identified with potential $A_m$.
The resulting BF equation $R_{mn} \equiv F_{mn} = 0$ is  different form the Maxwell equation
$\nabla_m F^{mn} = 0$. It is worth noting here that a similar difference
is between Maxwell and  Chern-Simons equations
in three dimensions.

The analysis of  equations \eqref{main} for  spin $s>1$ is more intricate because there are more
different rank fields and their gauge transformations including  both differential and  algebraic symmetry parameters.
All fields entering equations of motion  \eqref{main} can be divided into three subsets
consisting of dynamical fields, Stueckelberg fields, and auxiliary fields. \footnote{As usual, Stueckelberg fields
are assumed to be set to zero by virtue of their algebraic gauge transformations, while
auxiliary fields are expressed via dynamical fields. Despite the absence of local degrees of freedom
we keep using the term "dynamical fields" to designate independent fields of the theory.}
It turns out that such a
triple decomposition of the original field space can be  done in two different ways. In the following  sections
we explicitly demonstrate how it works in the spin-$2$ case, and then briefly  generalize
to arbitrary spins.

\subsection{Spin-$2$ example}

In the  spin $s=2$ case  fields $\omega_m^a$ and $\omega_m$   are identified with the zweibein  and Lorentz spin
connection (more precisely, with their fluctuations over $\ads$ values $h_m^a$ and $w_m$). Field equations
read off from \eqref{main} and \eqref{curva} at $s=2$
\be
\label{r}
R_{mn} \equiv \nabla_m \omega_n - \nabla_n \omega_m  -\Lambda\, h_m{}_{,b}\wedge  \omega_n^b +\Lambda\, h_n{}_{,b} \wedge\omega_m^b = 0\;,
\ee
and
\be
\label{ra}
R_{mn}^a \equiv \nabla_m \omega^a_n - \nabla_n \omega^a_m  + h_m^a \wedge\omega_n  -   h_n^a \wedge \omega_m = 0\;,
\ee
are invariant under  the gauge transformations
\be
\label{gt}
\delta \omega_m = \nabla_m \xi  -\Lambda h_m{}_{,c} \xi^c \;,
\qquad
\delta \omega_m^a = \nabla_m \xi^a + h_m^a \xi\;.
\ee
Using  dualized field redefinitions $\omega^a_m \equiv  \epsilon^{ab} e_{m,\,b}$ and
$\omega_m  \equiv  \epsilon_{ab}\omega_m^{[ab]}$, the  equations \eqref{r} and \eqref{ra} can be cast into the standard
form for zweibein  $e_m^a$ and Lorentz spin connection $\omega_{m}^{ab}$. In particular,
the above field redefinition succeeded by setting $\Lambda = 0$  reproduces the known  expressions
for curvatures of the $(1+1)$ Poincare algebra.

The standard choice of dynamical fields is achieved by using the gauge parameter $\xi$ as the Stueckelberg
one to gauge away the trace component of $\omega^a_m$. \footnote{Scalar parameter $\xi$ can be dualized
into an antisymmetric parameter $\xi^{[ab]}$. Obviously,
this choice corresponds to the standard gauge fixing of local
Lorentz symmetry.}
To this end, using the $\ads$ background frame $h^a_m$ one converts world indices into fiber ones  and then
decomposes $\omega^a_m$ into three $o(1,1)$ irreducible components  as follows
\be
\label{decomp}
\omega^{a|b} = \omega^{(ab)} +  \eta^{ab} \hat\omega +  \half \epsilon^{ab} \tilde{\omega}\;,
\ee
where $\eta_{mn}\omega^{(mn)}=0$.
Residual components of $\omega_m^a$ can be combined into a single traceful symmetric  tensor
$\varphi_{ab} = \epsilon_a{}^c \omega_{cb} +\half \eta_{ab} \tilde{\omega}$, so that its
traceless and trace parts are given by $\omega^{(ab)}$ and $\tilde{\omega}$.
Field $\varphi^{ab}$ is then identified as dynamical field with the gauge symmetry transformation
$\delta \varphi_{ab} = \nabla_a (\epsilon_b{}^c\xi_c) + \nabla_b (\epsilon_a{}^c\xi_c)$ (cf. our comments below
\eqref{gt}). Simultaneously,
one considers equation \eqref{ra} as the constraint which allows one to express auxiliary field $\omega_m$ via dynamical field
$\varphi^{ab}$.

Equation $\epsilon^{mn}R_{mn} = 0$ \eqref{r} is dynamical. In terms of $\varphi_{ab}$
it reads $\nabla^a \nabla^b \varphi_{ab}  - \frac{3}{2} \Box_{\ads} \varphi + 2\Lambda \varphi = 0$,
where $\Box_{\ads}$ is the wave operator for the  $\ads$ metric, and $\varphi = \eta^{ab}\varphi_{ab}$.
Obviously, this is the Jackiw-Teitelboim equation \eqref{JTeq} linearized around $\ads$ background.
Imposing the conformal gauge which  in our case says that the traceless part of $\varphi^{ab}$ is set to zero,
one is left with  $\varphi$ subjected to  the residual equation
\be
\label{Liouvilleliner}
\big(\Box_{\ads}  - 2\Lambda\big)\,\varphi = 0\;.
\ee
The above equation is invariant under the  residual  gauge transformation
$\delta \varphi = \epsilon_{ab}\nabla^a \xi^b $ with a parameter
satisfying the  Killing equation $\nabla^{a} \xi^{b}+\nabla^{b} \xi^{a} -\eta^{ab} \nabla_c\, \xi^{c} = 0$.
Therefore, one is left with the scalar field which is massive gauge field. \footnote{Recall here the well-known
example of a scalar gauge field given by the trace of the metric field.}

It is crucial that for  $\Lambda \neq 0$ the above choice of dynamical fields is not unique.
Another option is to use
the  gauge parameter $\xi^a$ as a Stueckelberg one to completely gauge away the field $\omega^m$ \eqref{gt}.
Then, from \eqref{r} and \eqref{decomp} one derives the  algebraic constraint $\Lambda \tilde{\omega} = 0$, which says that
the antisymmetric component of $\omega^a_m$ \eqref{decomp} is set to zero.
It follows that a dynamical field is again a traceful symmetric tensor  $\phi^{ab} = \omega^{(ab)} +  \eta^{ab} \hat\omega$, but now
its traceless and trace components are identified with
$\omega^{(ab)}$ and $\hat{\omega}$. The gauge transformations now read
$\delta\phi^{ab} = \Lambda^{-1}\nabla^a\nabla^b \xi + \eta^{ab}\xi$.
Note that it reproduces the gauge transformation of a spin-$2$ partially-massless field.

Equation $\epsilon^{mn}R^a_{mn} = 0$ \eqref{ra} is dynamical. Imposing the gauge $\eta_{ab}\phi^{ab} = 0$ one is left with a traceless component subjected to the
residual equation
\be
\label{conser}
\nabla^a J_{ab} = 0\;,
\ee
where $J^{ab} \equiv  \epsilon^{ac}\phi^{b}{}_c$ is  symmetric and traceless tensor field.
A residual  gauge invariance of \eqref{conser} is given by
$\delta J_{ab} = \Lambda^{-1}\epsilon_{ac} \nabla^c \nabla_b \xi+ \Lambda^{-1}\epsilon_{bc}\nabla^c \nabla_b \xi$
with the scalar parameter satisfying differential constraint $(\Box_{\ads}  - 2\Lambda)\xi = 0$. It follows
that equation \eqref{conser} can be seen as the conservation condition for a rank-$2$ current $J_{ab}$
on the  $\ads$ space.

\subsection{Standard realization }
\label{sec:sigmaplus}

By the standard realization  we mean using a set of dynamical fields  that generalize the metric-like  variable
of the spin-$2$ case. A careful analysis shows that all higher rank equations in \eqref{main} are constraints
except for the $k=0$ equation $\epsilon^{mn}R_{mn} = 0$ treated as
the dynamical equation.
The constraints allow one to express all fields via derivatives of dynamical fields which are
scalar and rank-$s$ traceless tensors, $\varphi$ and $ \varphi^{a_1 ... a_s}$. As in the
spin-$2$ case, the scalar component $\varphi$ is identified with antisymmetric component of $\omega_m^a$
\eqref{decomp}. The rank-$s$ traceless component $\varphi^{a_1 ... a_s}$ is identified with
a totally symmetrized highest rank $k=s-1$ connection
$\varphi^{ma_1 ... a_{s-1}} = \omega^{(m| a_1 ... \, a_{s-1})}$. Note that the symmetrized combination
is automatically traceless because all other components in $\omega_m^{a_1 ... a_{s-1}}$ are gauged away using the  Stueckelberg symmetry
transformations.

Quite analogously, one shows that the only independent gauge parameter is $\xi^{a_1 ... a_{s-1}}$,
while all others are treated as Stueckelberg parameters or auxiliary parameters expressed via derivatives of
$\xi^{a_1 ... a_{s-1}}$. The leftover gauge symmetry with parameter $\xi^{a_1 ... a_{s-1}}$
allows one to impose the higher spin gauge,
\be
\label{hsgc}
\varphi^{a_1 ...\, a_s} = 0\;,
\ee
which is legitimate since  any $o(1,1)$ rank-$k$ irreducible tensor for $k>0$ has
just two independent components. Then, the only dynamical field is given by the scalar $\varphi$.
Imposing the higher spin gauge \eqref{hsgc} and solving the constraints in \eqref{main} one finds that
the leftover equation reduces to the massive scalar equation with a particular value
of mass-like term
\be
\label{scaleq}
\big(\Box_{\ads}- m_s^2\big)\,\varphi  = 0\;,
\qquad \text{where}
\qquad
m_s^2  =  s(s-1)\Lambda\;,\quad s\geq 2\;.
\ee
This is invariant under the residual  gauge transformations,
\be
\label{leftgauge1}
\delta \varphi = \epsilon_{mn_1}\nabla^m \nabla_{n_2} \cdots \nabla_{n_{s-1}}\xi^{n_1 \cdots\, n_{s-1}}\;,
\ee
provided that the gauge parameter  satisfies   the generalized
Killing equation
$\nabla^{(a_1} \xi^{a_2 ... a_{s})} -\frac{1}{s-1} \;\eta^{(a_1a_2} \nabla_c\, \xi^{a_3 ... a_{s-1})c} = 0$,
which is simply the stability transformation for the higher spin gauge \eqref{hsgc}.
The   massive scalar field \eqref{scaleq} endowed with higher order  derivative transformations
for the scalar parameter \eqref{leftgauge1}  can be treated as topological partially-massless field of spin $s$
and the maximal depth.

\subsection{Dual realization}
\label{sec:sigmaminus}

By the dual realization  we mean using a set of dynamical fields  that generalize the current-like variable
of the spin-$2$ case. Quite analogously to the standard realization one shows that all higher rank equations
in \eqref{main} can be treated as constraints except for the  $k=s-1$ equation
$\epsilon^{mn}R_{mn}^{a_1 ... a_{s-1}} = 0$ which is dynamical.
The constraints allow one to express all fields via derivatives of dynamical fields also given by
scalar and rank-$s$ symmetric traceless tensor, $\phi$ and $\phi^{a_1 ... a_s}$. But now
the scalar  $\varphi$ is identified with a trace component of $\omega_m^a$ \eqref{decomp}, while the
rank-$s$ traceless field $\phi^{a_1 ... a_s}$ is identified with
the symmetric traceless  component of $k=s-1$ connection
$\omega_{m|a_1...a_{s-1}}$, namely $\phi_{ma_1 ... a_{s-1}}  = \omega_{(m|a_1...a_{s-1})} -traces$.  An independent gauge  parameter is the scalar  $\xi$, while the gauge transformations read
\be
\label{residualtr}
\delta \phi = \Box_{\ads}\xi  -s(s-1)\Lambda \xi\;,
\qquad
\delta \phi_{a_1 ... a_{s}} = \nabla_{a_1} \cdots  \nabla_{a_s} \xi + ...\;,
\ee
where the ellipsis refers to  proper symmetrizations and trace terms. Using gauge symmetry \eqref{residualtr} one can
impose the scalar gauge $\phi= 0$ along with the residual gauge parameter equation
$\Box_{\ads} \xi  -s(s-1) \Lambda  \xi =0$.

In the scalar gauge, the only dynamical field is the rank-$s$ component $\phi^{a_1 ... a_{s}}$.
Then, solving the constraints in \eqref{main} one finds that  the independent  dynamical equation is  given by
\be
\label{inteq}
\nabla^n J_{n a_1 ... a_{s-1}} = 0
\qquad \text{for} \qquad
J_{na_1 ... a_{s-1}} = \epsilon_n{}^m\phi_{ma_1 ... a_{s-1}}\;,
\ee
where dualized tensor field $J_{na_1 ... a_{s-1}}$ is totally symmetric and traceless.
One identifies $J_{a_1 ... a_{s}}$ with spin-$s$ conserved current on the $\ads$ space,
while higher order derivative transformations in \eqref{residualtr} are  "improvements".
On the other hand, the "improvement" transformation can treated as the gauge transformation of a
spin-$s$ partially massless field of maximal depth.

\section{Conclusions and outlooks}
\label{sec:conclu}

In this paper we have introduced $\ads$ higher spin gravity defined via topological BF
action with fields taking values in $sl(N, \mathbb{R})$ algebra. We identified
the spectrum of higher spin gauge field fluctuations over $\ads$ background, and for $\Lambda \neq 0$ found out
their dual description in terms of higher rank conserved currents.

Note that the proposed duality between two metric-like formulations guarantees that the two
realizations describe the same dynamical system. Indeed, in order to single out independent fields and parameters we used  Stueckelberg
algebraic symmetry and eliminated auxiliary fields of the  frame-like formulation. Whence, two metric-like
formulations \eqref{scaleq}
and \eqref{inteq} coming from the original (linearized) BF system are dynamically equivalent.
This type of duality is similar to that one in the WZNW theory with
the equations of motions for a group variable $g(x)$ given by a  second order field equation
$\d^m (g^{-1} \d_m g) = 0$. At the same time, introducing the current $J_m = g^{-1}\d_m g$ one equivalently
arrives at the  first order conservation condition $\d^m J_m = 0$ (\textit{e.g.}, see \cite{dual}).

It is instructive to compare linearizations of $\ads$ higher
spin theory for $\cA = sl(N, \mathbb{R})$ algebra and $3d$ Chern-Simons higher spin
theory for  $\cA \oplus \cA$ algebra. In the $3d$
case,  choosing $AdS_3$ spacetime as the background for higher spin
fields yields Fronsdal equations of motion for massless
gauge fields with spins $s = 2,..., N$ that describe,
however, no local degrees of freedom \cite{Henneaux:2010xg,Campoleoni:2010zq}.
In the $2d$ case, the same analysis does not yield
Fronsdal equations, although the system is also topological.
Instead, in the higher spin gauge which generalizes the conformal gauge in the gravity,
the dynamical equations in the standard realization are scalar field equations with higher order derivative
gauge symmetries guaranteeing   the absence of
propagating  modes. These gauge systems can be interpreted as two-dimensional topological partially-massless fields.

It is remarkable that BF higher spin theory considered in the
present paper has been in fact formulated within the unfolded
approach to higher spin dynamics. In  \cite{Alkalaev_soon} we
consider the theory from the unfolded formulation perspective and
use the cohomological technique to analyze BF equations of
motion. In particular, independent parameters, fields and
equations of the standard and the dual realizations elaborated in
sections \bref{sec:sigmaplus} and \bref{sec:sigmaminus} have a streamlined interpretation as  cohomology elements
of the so-called $\sigma_-$ and $\sigma_+$ nilpotent operators
acting in the field space. Also, using the  cohomological
machinery we analyze equations for 0-forms which were not
considered in the present paper.

The proposed BF higher spin theory can be extended so as to include an infinite tower of higher spin fields
provided a gauge  algebra $\cA$ is chosen to be one of infinite-dimensional higher spin algebras \cite{Fradkin:1989uh,Alkalaev_soon,Rey}.
The resulting theory is also topological. Recall that BF higher spin theory can however  have
local degrees of freedom provided that
fields take  values in particular  infinite-dimensional associative algebra \cite{Vasiliev:1995sv}.


\noindent \textbf{Acknowledgements.} I am grateful to  M. Grigoriev, E. Skvortsov and M.A. Vasiliev
for  useful comments and discussions. I would like to thank the Galileo
Galilei Institute for Theoretical Physics, Florence, Italy, for
hospitality during my staying at GGI Workshop on Higher Spins,
Strings and Duality, and to the Dynasty foundation for the financial support of
the visit. The work is supported by RFBR grant 12-02-31838.


\noindent \textbf{Note added.}
After the present paper  has been submitted to arXiv, we were informed by Soo-Jong Rey that $2d$ higher spin
theories were partly discussed in his talk  \cite{Rey}, mainly from the holographic perspective.

\providecommand{\href}[2]{#2}\begingroup\raggedright
\addtolength{\baselineskip}{-3pt} \addtolength{\parskip}{-1pt}

\end{document}